# A common origin for 3/4- and 2/3-power rules in metabolic scaling

Jinkui Zhao, SNS, Oak Ridge National Laboratory, Oak Ridge, TN 37831, USA,

Email:  zhaoj@ornl.gov

Fax:    865-241-5177

Phone: 865-574-0411

A central debate in biology has been the allometric scaling of metabolic rate. Kleiber's observation that animals' basal metabolic rate scales to the 3/4-power of body mass (Kleiber's rule) has been the prevailing hypothesis in the last eight decades. Increasingly, more evidences are supporting the alternative 2/3-power scaling rule, especially for smaller animals. The 2/3-rule dates back to before Kleiber's time and was thought to originate from the surface to volume relationship in Euclidean geometry. In this study, we show that both the 3/4- and 2/3-scaling rules have in fact one common origin. They are governed by animals' nutrient supply networks-their vascular systems that obey Murray's law. Murray's law describes the branching pattern of energy optimized vascular network under laminar flow. It is generally regarded as being closely followed by blood vessels. Our analysis agrees with experimental observations and recent numerical analyses that showed a curvature in metabolic scaling. When applied to metabolic data, our model accurately produces the observed 2/3-scaling rule for small animals of ~10 kg or less and the 3/4-rule for all animals excluding the smallest ones (~15 g). The model has broad implications to the ongoing debate. It proves that both the 3/4- and 2/3-exponents are phenomenological approximations of the same scaling rule within their applicable mass ranges, and that the 2/3-rule does not originate from the classical surface law.

\body **Introduction**

Allometric scaling laws are phenomenological observations that appear to be universal in governing the structure and function of animals of diverse body sizes. Despite many progresses, the underlying mechanisms of these laws are still poorly understood. One fundamental phenomenon is the scaling of metabolic rate ($B$) against animal's body mass ($M$). In 1932, Max Kleiber observed that a quarter power scaling relationship $B \sim M^{3/4}$ exists in animals from his study of 13 species ranging from dove to steer (1). Since then, Kleiber's rule has replaced the previously prevailing surface law as the predominant hypothesis. Surface law states that metabolism is limited by heat dissipation and thus scales to an animal's surface area. Hence $B \sim M^{2/3}$. Even though large amounts of experimental data support Kleiber's rule (2), recent studies have shown that the 2/3-power scaling rule cannot be ignored, especially for smaller animals of ~ 10 kg or less (3-6). Nonetheless, despite the fact the classical surface law yields a 2/3-power exponent, we will show in this work that the rule's origin lies elsewhere. Many recent efforts have been devoted to understanding the scaling rule, the most prominent of which is the fractal model proposed by West, Brown and Enquist (WBE) (7). WBE has had broad impacts to the field of animal allometry. However, like many pioneering theories, WBE is not without controversies (4,8,9). Several competing models have been proposed but no consensus has emerged (10-14).

Living organisms are complex entities. Their metabolisms are likely influenced by many factors. Nevertheless, animal physiology is essentially the same regardless of body sizes. Any fundamental rule that governs the metabolic rate can be expected to be applicable to animals large and small. One common measure of metabolism is the consumption of oxygen, which is transported through animals' vascular systems. It is then logical to assume that the ability of these resource distribution networks to deliver oxygen limits the metabolic rate. Such a consideration is entirely reasonable since maintaining blood incurs costs to animals (15). Energetically, it would be too expensive for animals to have redundant transport capacities such that $B$ is unhindered by the network. This network-limited principle is argued by WBE and followed by many other theories (10,12,13). In WBE, the vascular network is assumed to follow a pattern of area preserving fractals (7). However, in 1926, Cecil Murray applied the principle of minimum energy cost and showed that the bifurcation of vascular vessels under laminar flow follows (15): $r_p^3 = r_{d1}^3 + r_{d2}^3$. $r_p$, $r_{d1}$ and $r_{d2}$ are the radii of the parent vessel and its two daughter branches respectively. For symmetric networks, Murray's branching law can be expressed as

$$r_p^3 = n r_d^3 \qquad (1)$$

$n$ is the number of daughter branches. The majority of blood vessels, apart from aorta and large arteries, obey Murray's law (16-20). We therefore expect Murray's law to play a crucial role in animals' metabolism and metabolic allometry.

**Model**

We propose that animals' vascular systems that obey Murray's law (Murray's networks) form the underlying foundation for metabolic allometry. We show that the resulting scaling rule from such networks is in perfect agreement with experimental data for both small and large animals. Our general approach for obtaining oxygen consumption and hence the metabolic rate is to first calculate the resistance and the blood flow rate through the vascular network. For these calculations, we make three assumptions:

    A1: Animal's vascular networks are spacing filling.

    A2: The structures of the network's terminal branches, the capillaries, are invariant (7,21).

    A3: The blood pressure $\Delta P$ across the whole vascular system is independent of body mass $M$ (7,21,22). In the same time, the oxygen pressure in blood scales with $M^{-1/12}$ (23,24).

Additionally, the usage of Murray's networks implies energy minimization for the vascular system. Assumption A1 is a logical consequence of resource-delivering networks. Such a network has to branch into every part of an animal's body to deliver the needed nutrients. A1 and A2 are essentially the same as those in WBE, with one important distinction: In the current study, we make no statements on the blood pressure across a single capillary. Consequently, we do not assume an invariant flow rate or a constant metabolic rate through individual capillaries. Rather, we use assumption A3 to derive the blood flow rate through the vascular system. Animal's blood pressure is generally observed to be constant regardless of body mass (7,21,22). On the one hand, sufficient blood pressure is required to overcome the gravitational and viscous resistances of the blood (22). On the other hand, even though higher pressures should allow for faster blood circulation and benefit metabolism, animals' heart and arterial tissues have limited strength. Higher pressures will require significantly stronger tissues, as can be seen in the case of Giraffes. In order to deliver bloods to their heads, Giraffes' blood pressure is about twice that of other animals. To produce and withstand even such a nominally increased pressure, Giraffes' arterial systems have exceptional thick walls (22).

It has been observed that oxygen pressure in blood scales to animal body mass with $M^{-1/12}$ (23,24). Such variation is related to oxygen affinity of the hemoglobin protein (22) and appears to be a size dependent evolutionary adaptation in animals (25). Because metabolism is proportional to oxygen consumption, we therefore take blood oxygen pressure into account in assumption A3: the metabolic rate is proportional to both the rate of blood flow and the oxygen pressure in the blood.

In the following, we establish the relationship between the flow resistance $Z$ of a Murray's network and an animal's body mass $M$. A simplified vascular system as a linearly branching network is shown in figure 1. Each level of successive branching is represented by a rank index $k$. To calculate $Z$, we first derive the total number of ranks as a function of mass $M$. We then prove that under the space filling assumption A1, Murray's network is impedance matched. Namely, the combined resistance is unchanged from one rank of the branches to the next.

At an arbitrary rank $k$, the total number of blood vessels is given by $N_k = n_1 n_2 \ldots n_k$ (Fig. 1). For simplicity, $N_k$ is rewritten as $N_k = n^k$. The constant $n$ can be regarded as the average degree of branching from rank 1 to $k$. From the space filling assumption A1, the volume $v_k$ served by a blood vessel of length $l_k$ and radius $r_k$ ($r_k \ll l_k$) corresponds to the space spanned by the vessel, which can be approximated by a sphere (7), or simply by a cube with $v_k = l_k^3$. At the terminal capillary branches of the network (rank $k_c$, the subscript $c$ denote for capillary), the total service volume $V_c$ of all the capillaries must equal to animal's body volume: $V_c = M/\rho$. $\rho$ is animal's mass density ($\rho \approx 1$ g/cm$^3$). In total, there are $N_c = n^{k_c}$ capillaries, each with a service volume of $v_c = l_c^3$ and a service mass of $m_c = \rho v_c$. We therefore have $M = N_c m_c = n^{k_c} m_c$. From assumption A2, $l_c$, $v_c$, and $m_c$ are invariant regarding to body mass $M$. Thus, the number of capillaries $N_c$ scales linearly with body mass $M$ and the capillary number density remains constant: $N_c = n^{k_c} = M/m_c$. From this relationship, the total number of ranks in the vascular network is given by the capillary rank index $k_c$:

$$k_c = \log(M/m_c)/\log(n) \qquad (2)$$

We now show that a space filling Murray's network leads to impedance matching. First, we demonstrate that the flow resistance $z_k$ of a parent vessel at rank $k$ is the same as the combined values $Z'_{k+1}$ of all its $n_{k+1}$ daughter vessels at rank $k+1$. For a single blood vessel with radius $r_k$ and length $l_k$, $z_k$ is given by Poiseuille's law: $z_k = (8\eta\pi^{-1})l_k r_k^{-4}$. $\eta$ is blood viscosity. It follows that $z_k/z_{k+1} = (l_k/l_{k+1})(r_k/r_{k+1})^{-4}$. The ratio of the radii is readily given by equation 1 (with $p = k$ and $d = k+1$): $r_k/r_{k+1} = n_{k+1}^{1/3}$. To obtain $l_k/l_{k+1}$, we go back to the space filling assumption A1: A rank $k$ vessel and its $n_{k+1}$ daughter branches serve the same volume (7). Namely, $v_k = n_{k+1}v_{k+1}$. Hence, $l_k/l_{k+1} = n_{k+1}^{1/3}$. Combining these relations together yields $z_k = z_{k+1}/n_{k+1}$. Because blood flows through the $n_{k+1}$ daughter branches in parallel, the collective flow resistance of these daughter branches is then given by $Z'_{k+1} = z_{k+1}/n_{k+1}$, which is the same as the resistance $z_k$ of the parent vessel.

Next, with the same argument as above, the collective flow resistance of all $N_k$ vessels at rank $k$ is given by $Z_k = z_k/N_k$. By comparing $Z_k$ with $Z_{k+1}$ and noting that $N_{k+1} = n_{k+1}N_k$ (Fig. 1), we have $Z_k/Z_{k+1} = (z_k/z_{k+1})(N_{k+1}/N_k) = 1$. Thus, the flow resistance is conserved between generations of branching: $Z_0 = Z_1 = \ldots = Z_c$. Impedance matching is an important characteristic of optimized electrical circuitry and transport pipes, including vascular systems (26). The above analysis shows that a spacing filling Murray's network indeed possesses this optimized feature.

Since the branching generations are connected in series, the overall flow resistance of a space filling Murray's network is given by $Z = Z_0 + Z_1 + \ldots + Z_c = k_c Z_c$. The resistance $Z_c$ of all $N_c$ capillaries at rank $k_c$ is $Z_c = z_c/N_c = z_c m_c/M$. At last, we arrive at the flow resistance of a space filling Murray's network:

$$Z = k_c Z_c = k_c z_c m_c/M. \tag{3}$$

In animals such as mammals and birds, the aorta and main arteries are optimized for pulsatile and turbulent flows. Transition to laminar flow takes place a few ranks away from the aorta (7). In the current discussion, we can view the aorta and the main arteries, which are elastic, as pressure buffers for the pulsating heart. They take in blood during the rather short period of the cardiac contraction, and then exert a constant pressure onto the arterioles and capillaries. This is not an unreasonable simplification since blood pressures are typically measured at the main arteries rather than at the heart itself. Under such a consideration and assuming that the transition from pulsatile to laminar flows occurs at rank $\bar{k}$, the effective resistance of the vascular network is now modified to $Z = (k_c - \bar{k})z_c m_c/M$. Replacing $k_c$ with equation 2 and noting the properties of logarithmic functions, $\log(n^{\bar{k}}) = \bar{k}\log(n)$ and $\log(M/m_c) - \log(n^{\bar{k}}) = \log(M/m_c/n^{\bar{k}})$, we have

$$Z = [z_c m_c/\log(n)]\cdot\log(M/\beta)/M,$$
$$\text{with } \beta = m_c n^{\bar{k}} \tag{4}$$

$\bar{k}$ corresponds to the same quantity as in WBE (7). It's possible that the aorta and main arteries might play a more significant role in metabolic scaling, especially in larger animals. However, aside from a modified $Z$, taking $\bar{k}$ into account does not affect the basic formalism of our model in the following.

With the flow resistance $Z$ derived in equation 4, we now calculate the blood flow rate $Q$, oxygen flow rate $Q_O$, and the metabolic rate $B$. According to Poiseuille's law, the mass flow rate $Q$ of blood through a network with a resistance $Z$ under laminar flow is given by $Q = \Delta P/Z$. From assumption A3, blood pressure $\Delta P$ is constant. Using equation 4, we obtain $Q = [\Delta P \log(n)/(z_c m_c)] \cdot M/\log(M/\beta)$. Also from assumption A3, the partial pressure of blood oxygen scales with $M^{-1/12}$. Therefore, the oxygen flow rate $Q_O$ is given by $Q_O = \gamma M^{-1/12} Q$, $\gamma$ is a proportionality constant. Finally, the metabolic rate is proportional to oxygen consumption, $B = \eta Q_O$, $\eta$ is a proportionality constant. Combining the expressions for $B$, $Q_O$, $Q$, and $Z$ together, we have:

$$B = \alpha M^x/\log(M/\beta),$$
$$\text{with } x = 11/12 \text{ and } \alpha = \gamma \eta \Delta P \log(n)/(z_c m_c) \tag{5}$$

The constant $\beta$ is the same as defined in equation 4. Both constants $\alpha$ and $\beta$ are related to the properties of the vascular system. The constant $\beta$, for example, reflects both the volume serviced by a single capillary and the branching structure of the network.

**Results**

Equation 5 describes the metabolic rate as a function of body mass for animals with vascular systems that obey Murray's law. We now apply our model to the basal metabolic data compiled by McNab (27,28) and validate our approach. The first test is to fit equation 5 (with fixed $x = 11/12$ value) to a small subset of data with somewhat arbitrary $M \leq 1000$ g. Then, using the obtained coefficients $\alpha$ and $\beta$, we examine the calculated $B$ values and compare them to the metabolic data for larger animals. As evident from figure 2, the calculation describes the metabolic rate for all animals remarkably well, despite the fact the coefficients $\alpha$ and $\beta$ were obtained from small animals only.

The $M^{-1/12}$ scaling of blood oxygen pressure is an empirical observation (23) and is not the result of a rigorous theory. For this reason, we relax the fitting condition and allow $x$ in equation 5 to vary. Using the whole data set, we obtain $x = 0.868\pm0.013$, $\alpha = 0.023\pm0.004$ and $\beta = 0.34\pm0.17$. This $x$ value is slightly less than 11/12. To understand the meaning of the $\beta$ value, we take $n = 3$ and $\bar{k} = 7$ as discussed by WBE (7) as an example. From equation 4, $\beta = m_c n^{\bar{k}}$. $\beta = 0.34$ then gives a capillary service volume of $v_c \sim 0.16$ mm$^3$ (with $v_c = m_c/\rho$ and $\rho \approx 1$ g/cm$^3$) and a capillary length of $l_c \sim 0.5$mm. This result is surprisingly close to experimental observations (18,30), lending further supports to our analysis.

As a result of logarithmic normalization, equation 5 has a convex curvature on the logarithmic scales. Such a curvature in the metabolic data has been observed recently (28,26,29). The logarithmic normalization itself is an intrinsic property of linearly branching networks. We can no longer speak of a scaling exponent. However, as we see from figure 2, the curvature of equation 5 is rather gentle. Therefore, the slope of $\log(B)$ vs. $\log(M)$ (slope $y = x - 1/\ln(M/\beta)$, ln is the natural logarithm) can be viewed as a mass dependent scaling exponent. From the fitting result in the last paragraph, the average value of slope within the limits of the used experimental data ($M = 2.2$ g to $3.22\times10^6$ g) is $y = 0.716$. However, if we exclude the smallest animals, we obtain an average value of $y = 0.75$ for $M = 15$ g to $3.22\times10^6$ g (Fig. 2), which is precisely Kleiber's 3/4-rule. Thus, the 3/4-power scaling rule is valid when all except the smallest animals are considered. Namely for $M \geq 15$ g (1,2). On the other hand, the average slope from $M = 2.2$ g to $10^4$ g is $y = 0.665$, which is the 2/3-power rule. This shows that the mechanism for the observed 2/3-power rule for small animals ($M \leq 10^4$ g) (3-6) is the same as that for Kleiber's rule. Both Kleiber's and the 2/3-power rules are linear approximations of equation 5 within the above said body mass boundaries.

Shown in figure 3, animals' field metabolic rates (FMR) compiled by Capellini et al. (6) exhibit a similar curvature as the basal metabolic rates. These FMR data are equally well described by equation 5, providing evidence that surface heat dissipation plays little role to animals' FMR. Incidentally, these FMR data include both placental mammals and marsupials. For marsupials, allowing $x$ in equation 5 to vary results in a lower value of $x = 0.75$ (Fig. 3). Interestingly, marsupials' dependence of blood oxygen pressure on body mass has a ~3× steeper slope than that of placental mammals (31). If we consider this difference together with the $M^{-1/12}$ oxygen pressure scaling used for other mammals, we expect marsupials' blood oxygen pressure to scale with $M^{-3/12}$ and $x$ to be 3/4, which is the exact result from the above analysis. Admittedly, there is a high degree of uncertainty here due to the limited availability of experimental data. The reported slope in (31) is also small in absolute term. However, what is demonstrated here is that metabolic rate is in fact limited by the vascular system's ability to deliver oxygen: Comparing to placental mammals, marsupials' oxygen blood pressure decreases faster with body mass $M$ as $M$ increases. Consequently, the rate of increase in metabolism for larger marsupials is slower than that for larger placental mammals. Namely, marsupials must have a smaller slope value in their metabolic scaling.

**Discussion**

The model for animals' metabolic allometry presented in the current study is based on the well established Murray's vascular network. As mentioned earlier, one key difference in our approach from the seminal WBE is that we assume invariant blood pressure across the whole network, not across single capillaries. In fact, because the flow resistances of successive ranks in Murray's network are matched, the pressure drop across a single capillary is simply given by $\Delta P/k_c$, which varies with body mass (see eq. 2). Thus, the metabolic rate of a single capillary in our model is not an invariant.

As discussed in the Result section, our model predicts that Kleiber's 3/4-rule applies to larger animals with $M \geq 15$ g and the 2/3-rule applies to smaller ones with $M \leq 10^4$ g, agreeing with experimental observations (1-6). One factor that has fueled the seemly irreconcilable dispute over which of these scaling rules is valid lies with the fact that these boundaries are not very sharply defined. The curvature of equation 5 is rather gentle and there is a high degree of scatter in the experimental data (see Fig. 2). As the slope of $\log(B)$ vs. $\log(M)$ from equation 5 is given by $y = x-1/\ln(M/\beta)$, we see that animals with larger $M$ must have a steeper slope value $y$. Conceptually, this can be approximated as the following. On the one hand, the combined flow resistance of all the capillaries $Z_c$ decreases with $M$ since there are more capillaries for larger animals and these capillaries work in parallel (see Eq. 3). On the other hand, the number of ranks in the branching network $k_c$ increases with $\log(M)$ (Eq. 2). Therefore, in a simplified view, the total resistance $Z$ increases with $\log(M)$ and decreases with $M$ at the same time. Since $\log(M)$ is a slower increasing function than $M$, the resistance $Z$ will thus have a faster rate of decrease at larger $M$ values. Therefore, the metabolic rate for larger animals should have a steeper slope than that for smaller ones. From equation 5, the curvature of metabolic scaling is determined by $\beta$. We obtain the $\beta$ value through the application of equation 5 to experimental metabolic data.

However, from the point of view of the vascular network, β is determined by the properties of the vascular network, $l_c$, $n$, and $\bar{k}$ (Eq. 4). Therefore, the curvature and the boundaries of where the transition from the 2/3- to the 3/4-rules occur are a property of the vascular network. They are determined by the capillary size $l_c$, the degree of branching $n$, and the transition rank $\bar{k}$ from pulsatile to laminar flows.

Murray's network is not area preserving. Higher rank branches have larger total cross sections. As has been pointed out in (7), this fact can be used to explain the phenomenon of blood slowing in the capillaries. Using equation 1, the relationship between the radii of the aorta $r_0$ at rank 0 and of the capillaries $r_c$ at rank $k_c$ is readily established: $r_0/r_c = N_c^{1/3}$. At any rank $k$, the blood speed $u_k$ is inversely proportional to the rank's total cross section $S_k$, which is given by $S_k = N_k \pi r_k^2$. Consequently, the blood speeds in the aorta $u_0$ and in the capillaries $u_c$ are related by $u_0/u_c = S_c/S_0 = N_c r_c^2/r_0^2 = N_c^{1/3}$. For $N_c \sim 10^{10}$ as in humans, this gives an expected $\sim 10^3$ times slower blood speed in the capillaries than in the aorta, agreeing with observations (7).

From equation 5, we see that our model does not allow for animals with mass values of $M \leq \beta$. To understand this limitation, let's first ignore the pulsatile to laminar flow transition: With $\bar{k} = 0$, we have $\beta = m_c$. (Eq. 4). Therefore, β is the mass serviced by a single capillary. Obviously, in vascular networks where the base units, the capillaries are invariant, the animal mass has to be larger than that of $m_c$. In another word, there must be more than one capillary in the vascular system. When $\bar{k}$ is considered, the minimum allowed $M$ value by the model should correspondingly increase with $\bar{k}$. This is because $M$ should be sufficiently large such that there are enough branches in the vascular network to allow a pulsatile to laminar flow transition to actually occur.

Our assumptions A1 and A2 mean that the number of capillaries $N_c$ in our model scales linearly with body mass $M$. We note that there are arguments for variable capillary dimensions and number densities (23). The variations advocated in (23) state that both capillary's dimension and number density vary with body mass $M$, but the total capillary volume scales linearly with $M$: $N_c r_c^2 l_c \propto M$. We can show that such variations do not change our result on the metabolic rate $B$. Under the space filling assumption A1 ($N_c l_c^3 \propto M$), such variations lead to $l_c \propto r_c$. Recall that the flow resistance for a single capillary is $z_c \sim l_c r_c^{-4}$ and for all the capillaries is $Z_c = z_c/N_c$, we obtain $Z_c \sim l_c^4 r_c^{-4}/M \sim M^{-1}$ (compare to Eq. 3). Thus, the combined resistance of the capillaries $Z_c$ remains unchanged with respect to such capillary variations. We can then follow the same discussion to derive equations 4 and 5.

Lastly, as mentioned earlier, the treatment of the blood flow and pressure in the aorta and the main arteries does not affect the basic formalism of our model. However, with the $\bar{k}$ value taken into account, our model yields a capillary dimension from the metabolic data that agrees very well with the direct measurements on the capillaries. In (7), WBE argued that the pulsatile to laminar transition rank $\bar{k}$ varies with $M$: $\bar{k} \propto \log(M)$. If we replace $\bar{k}$ with $\bar{k} = k'\log(M)$, it is readily shown that, with modified constants $\alpha$ and $\beta$ ($\beta$ now defined by $\log(\beta) = \log(m_c)/[1-k'\log(n)]$), the form of equation 5 remains unchanged.

In conclusion, by employing Murray's vascular network, our model reveals the underlying foundation for animals' metabolic scaling. It explains the different scaling behaviors of basal metabolic rates as rooted in animals' vascular system whose branching pattern is governed by Murray's law. The 2/3-power scaling rule, valid for $M \leq 10^4$, is not a result of the surface law as has been generally assumed. It has the same mechanistic origin as that of Kleiber's 3/4-scaling rule, which is applicable for $M \geq 15$ g. Our model applies to field metabolic rates as well, suggesting that surface heat dissipation plays little or no role in animal's FMR.

**Acknowledgements**: This manuscript has been authored by UT-Battelle, LLC, under Contract No. DE-AC05-00OR22725 with the U.S. Department of Energy.

## References

1. Kleiber M (1932) Body size and metabolism. *Hilgardia* 6:315-353.
2. Savage VM, Gillooly JF, Woodruff WH, West GB, Allen AP, Enquist BJ, Brown JH (2004) The predominance of quarter-power scaling in biology. *Funct. Ecol.* 18:257-282.
3. Dodds PS, Rothman DH, Weitz JS, (2001) Re-examination of the "3/4-law" of metabolism. *J. Theor. Biol.* 209:9-27.
4. Kozlowski J, Konarzewski M (2005) West, Brown and Enquist's model of allometric scaling again: the same questions remain, *Func. Ecol.* 19:739-743.
5. White CR, Seymour, RS (2003) Mammalian basal metabolic rate is proportional to body mass$^{2/3}$. *Proc. Natl Acad. Sci. USA* 100:4046–4049.
6. Capellini I, Venditti C, Barton RA (2010) Phylogeny and metabolic scaling in mammals. *Ecology* 91:2783-2793.
7. West GB, Brown JH, Enquist BJ (1997) A general model for the origin of allometric scaling laws in biology. *Science* 276:122-126.
8. Kozlowski J, Konarzewski M, (2004) Is West, Brown and Enquist's model of allometric scaling mathematically correct and biologically relevant? *Func. Ecol.* 18:283-289.
9. Chaui-Berlinck JG (2006) A critical understanding of the fractal model of metabolic scaling, *J. Exp. Biol.* 209:3045-3054.
10. Banavar JR, Maritan A, Rinaldo (1999) A Size and form in efficient transport networks. *Nature* 399:130-134.
11. Darveau CA, Suarez RK, Andrews RD, Hochachka PW (2002) Allometric cascade as a unifying principle of body mass effects on metabolism. *Nature* 417:166-170.
12. Banavar JR, Moses ME, Brown JH, Damuth J, Rinaldo A, Sibly RM, Maritan A (2010) A general basis for quarter-power scaling in animals. *Proc. Natl Acad. Sci. USA* 107:15816-15820.
13. Dodds P, (2010) Optimal Form of Branching Supply and Collection Networks. *Phys. Rev. Lett.* 104, 048702.
14. Glazier DS (2010) A unifying explanation for diverse metabolic scaling in animals and plants. *Biol. Rev.* 85:111-138.
15. Murray CD (1926) The physiological principle of minimum work. I. The vascular system and the cost of blood volume. *Proc. Natl Acad. Sci. USA* 12:207-214.

16. Sherman TF (1981) On connecting large vessels to small, the meaning of Murray's law. *J. Gen. Physiol.* 78:431-453.
17. Labarbera M (1990) Principles of Design of Fluid Transport Systems in Zoology. *Science* 249:992-1000.
18. Nordsletten DA, Blackett S, Bentley MD, Ritman EL, Smith NP (2006) Structural morphology of renal vasculature. *Am J Physiol Heart Circ Physiol.* 291:H296-309.
19. Zamir M, (1988) Distributing and delivering vessels of the human heart. *J. Gen. Physiol.* 91:725-7365.
20. Kassab GS (2006) Scaling laws of vascular trees: of form and function. *Am. J. Physiol. Heart. Circ. Physiol.* 290:H894-903.
21. Schmidt-Nielsen K (1984) *Scaling: Why Is Animal Size so Important?* (Cambridge Univ. Press, Cambridge, MA).
22. Schmidt-Nielsen K (1997) *Animal physiology: adaptation and environment.* (Cambridge Univ. Press, Cambridge, MA).
23. Dawson TH (2005) Modeling of vascular networks. *J. Exp. Biol.* 208:1687-1694.
24. Schmidt-Neilsen K, Larimer JL (1958) Oxygen dissociation curves of mammalian blood in relation to body size. *Am. J. Physiol.* 195:424-428.
25. Singer D (2004) Metabolic adaptation to hypoxia: cost and benefit of being small. *Resp. Physiol. & Neurobiol.* 141:215-228.
26. Savage VM, Deeds EJ, Fontana W (2008) Sizing Up Allometric Scaling Theory. PLoS 4(9): e1000171.
27. McNab BK (2008), An analysis of the factors that influence the level and scaling of mammalian BMR. *Comp. Biochem. Phys. A* 151:5-28.
28. Kolokotrones T, Savage V, Deeds EJ, Fontana W (2010) Curvature in metabolic scaling. *Nature* 464:753-757.
29. Clarke A, Rothery P, Isaac NJB (2010) Scaling of basal metabolic rate with body mass and temperature in mammals. *J. Anim. Ecol.* 79(3) 610-619.
30. Staub NC, Schultz EL (1968) Pulmonary Capillary length in dog cat and rabbit. *Resp. Physiol.* 5:371-378.
31. Bland DK, Holland RAB (1977) Oxygen affinity and 2,3-diphosphoglycerate in blood of Australian marsupials of differing body size. *Resp. Physiol.* 31:279-290.

**Figure Captions**

**Figure** 1. Left: A segment of a simplified, symmetric vascular network. Branching rank index starts at the aorta with $k = 0$ and ends with the capillaries with $k = k_c$. Throughout this work, lower case letters are used to denote properties of a single blood vessel. Capital letters represent combined properties within a rank. $n_k$ is the degree of branching at rank $k$. $N_k$ is the total number of branches at rank $k$ and satisfies $N_{k+1} = n_{k+1} N_k$. Right: Parameter for a rank $k$ tube with length $l_k$ and radius $r_k$. The condition of $r_k \ll l_k$ is assumed for all vessels in the current discussion.

**Figure** 2. Scaling of basal metabolic rate. The circles are data compiled by McNab (27). We use the Levenberg-Marquardt algorithm (http://www.netlib.org/minpack/) to perform least-square fittings. The solid red line is the fitting of equation 5 to $M \leq 1000$ g data with $x = 11/12$ ($\alpha = 0.0162 \pm 0.0008$, $\beta = 0.66 \pm 0.12$). The dashed green line is the fitting result to all the data with a variable $x$ ($x = 0.868 \pm 0.013$, $\alpha = 0.023 \pm 0.004$ and $\beta = 0.34 \pm 0.17$). The average slope of the green line is 0.665 for M ≤ 10000 g (indicated by the dotted orange line below data points) and 0.75 for M ≥ 15 g (dotted purple line above data points). The arrows indicate the approximate limits where 3/4- and 2/3-scaling rules are valid.

**Figure** 3. Scaling of field metabolic rate. The FMR data are converted to watt from Capellini et al. (6). The solid red line is the least-square fitting of equation 5 to placental mammals ($\alpha = 0.044 \pm 0.005$, $\beta = 1.1 \pm 0.5$). The dashed green line is the fit to all mammal data ($\alpha = 0.034 \pm 0.003$, $\beta = 2.2 \pm 0.5$). Both fits use fixed value of $x = 11/12$. If $x$ is allowed to vary, we obtain $x = 0.933 \pm 0.026$ for all mammals, which is close to 11/12. However, for marsupials, we obtain $x = 0.751 \pm 0.057$ (dotted blue line). This low $x$ value could be explained by the fact that marsupials' blood oxygen pressure scales to body mass with a much steeper slope than other mammals (31).

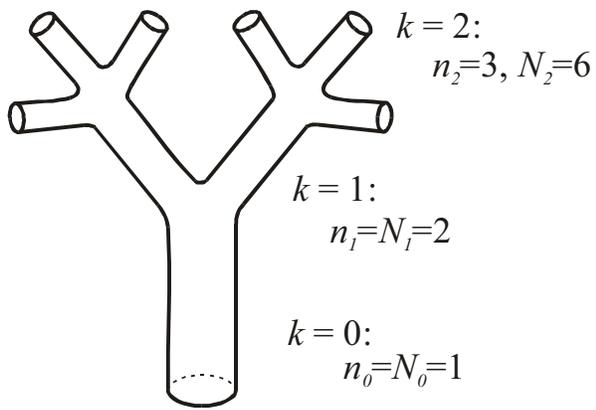 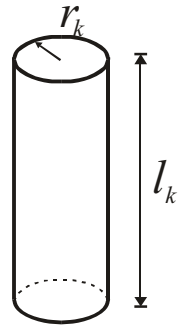

*Rank k = 0 1 2 ... $k_c$*

k = 2:
  $n_2$=3, $N_2$=6

k = 1:
  $n_1$=$N_1$=2

k = 0:
  $n_0$=$N_0$=1

$r_k$

$l_k$

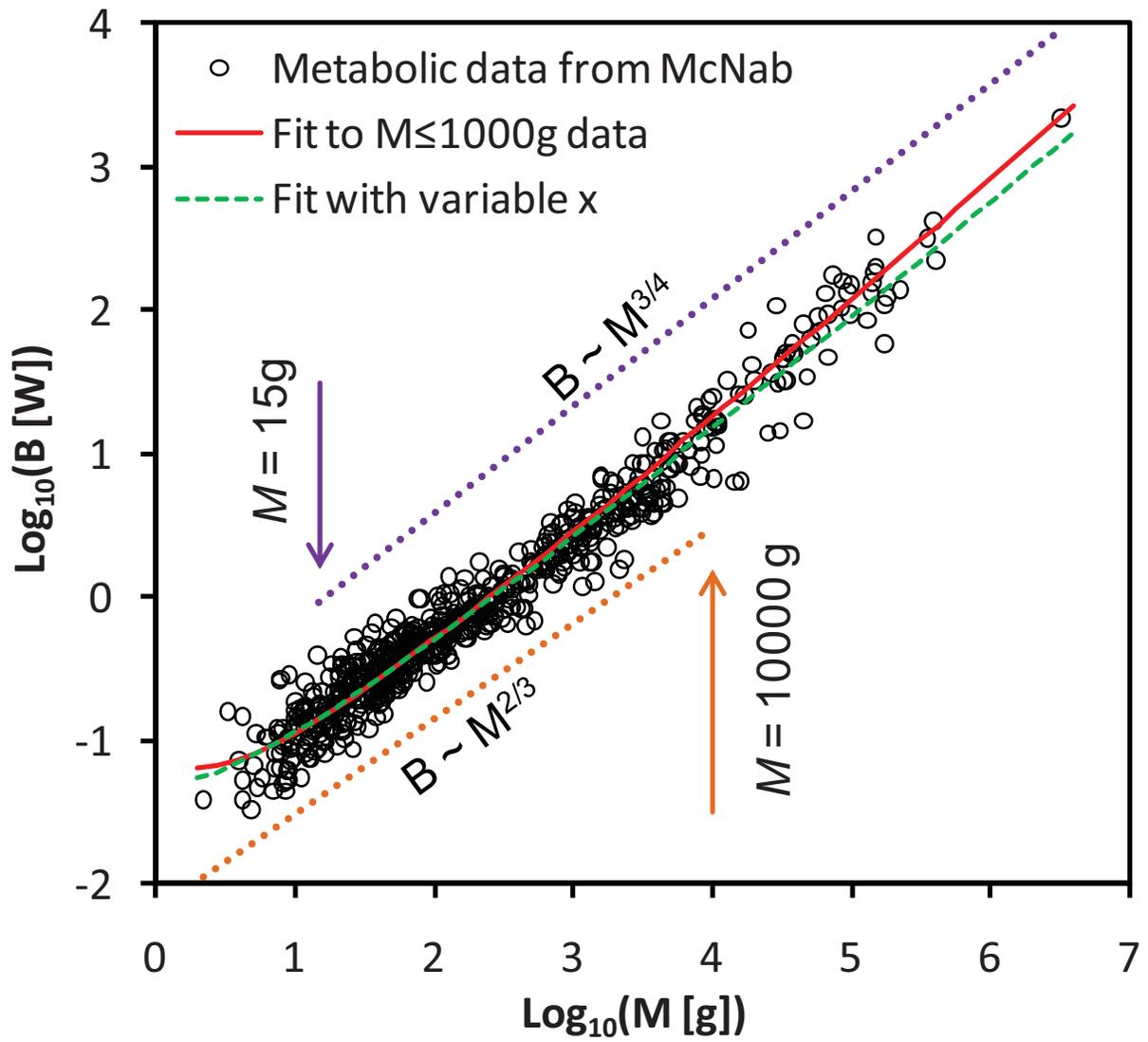

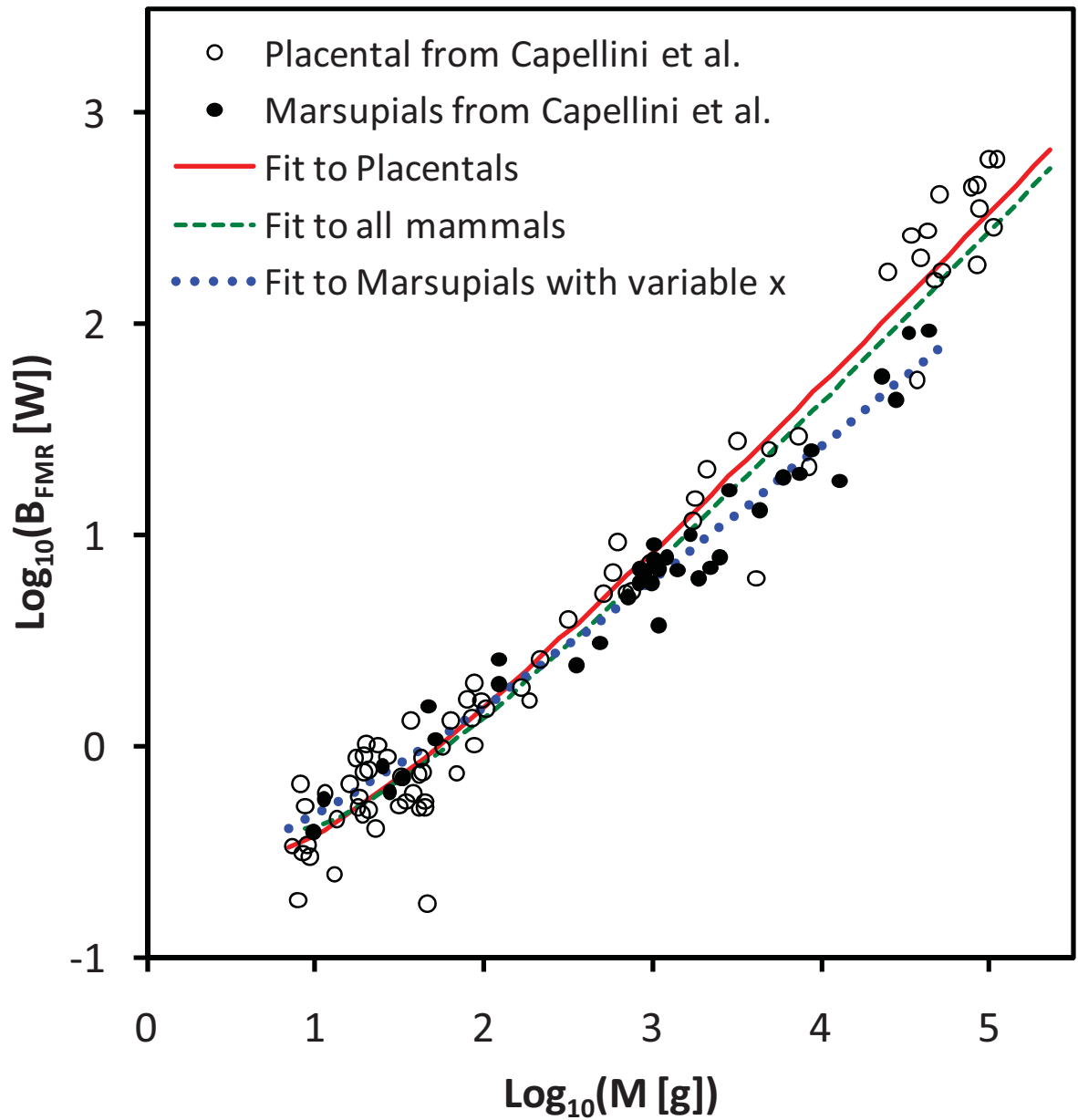